# Study of circular cross-section plasmas in HL-2A tokamak: MHD equilibrium, stability and operational $\beta$ limit[*]

SHEN Yong[1], DONG Jiaqi[1], SHI Zhongbing[1], HE Hongda[1], ZHAO Kaijun[2], PENG Xiaodong[1], QU Hongpeng[1], LI Jia[3], SUN Aiping[1]

1. Southwestern Institute of Physics, Chengdu 610041, China

2. College of Nuclear Science and Engineering, East China University of Technology, Nanchang 330013, China

3. School of Mathematics and Science, Chengdu University of Technology, Chengdu 610059, China

**Abstract**

Circular cross-section plasma is the most basic form of tokamak plasma and the fundamental configuration for magnetic confinement fusion experiments. Based on the HL-2A limiter discharge experiments, the magnetohydrodynamic (MHD) equilibrium and MHD instability of circular cross-section tokamak plasmas are investigated in this work. The results show that when $q_0 = 0.95$, the internal kink mode of $m/n = 1/1$ is always unstable. The increase in plasma $\beta$ (the ratio of thermal pressure to magnetic pressure) can lead to the appearance of external kink modes. The combination of axial safety factor $q_0$ and edge safety factor $q_a$ determines the equilibrium configuration of the plasma and also affects the MHD stability of the equilibrium, but its growth rate is also related to the size of $\beta$. Under the condition of $q_a > 2$ and $q_0$ slightly greater than 1, the internal kink mode and surface kink mode can be easily stabilized. However the plasma becomes unstable again and the instability intensity increases as $q_0$ continues to increase when $q_0$ exceeds 1. As the poloidal beta ($\beta_p$) increases, the MHD instability develops, the equilibrium configuration of MHD elongates laterally, and the Shafranov displacement increases, which in turn has the effect on suppressing instability. Calculations have shown that the maximum $\beta$ value imposed by the ideal MHD

---

mode in a plasma with free boundary in tokamak experiments is proportional to the normalized current $I_\mathrm{N}$ ($I_\mathrm{N} = I_\mathrm{p}(\mathrm{MA})/a(\mathrm{m})B_0(\mathrm{T})$), and the achievable maximum beta $\beta(\max)$ is calibrated to be $2.01 I_\mathrm{N}$, i.e. $\beta(\max) \sim 2.01 I_\mathrm{N}$. The operational $\beta$ limit of HL-2A circular cross-section plasma is approximately $\beta_\mathrm{N}^\mathrm{c} \approx 2.0$. Too high a value of $q_0$ is not conducive to MHD stability and leads the $\beta$ limit value to decrease. When $q_0 = 1.3$, we obtain a maximum value of $\beta_\mathrm{N}$ of approximately 1.8. Finally, based on the existing circular cross-section plasma, some key factors affecting the operational $\beta$ and the relationship between the achievable high $\beta$ limit and the calculated ideal $\beta$ limit are discussed.

**1. Introduction**

The magnetohydrodynamic (MHD) instability limits the efficiency of tokamak reactors[1,2]. Reactor efficiency is generally measured by plasma beta ($\beta$), and $\beta$ is the ratio of average plasma pressure to average magnetic pressure. Because the range of operating parameters, including the maximum operating beta of the plant, is limited when macroscopic instability occurs[3]. To optimize the plasma beta based on the study of the beta limit or the measurement of the beta limit[4] is the basic premise to realize the high beta operation and optimize the design of the device.

Benchmark tokamaks generally have a high-aspect-ratio configuration with circular cross-section and $\beta \sim \varepsilon^2$, where $\varepsilon$ stands for inverse aspect ratio. By studying the circular cross-section plasma, we can find the basic properties of tokamak, and facilitate the modification of the device and plasma profile in various forms in the future[5]. The HL-2A tokamak is a medium-sized tokamak with a two-zero closed divertor, which can perform limiter discharge in a circular configuration and also realize divertor discharge, thus obtaining a nearly circular divertor configuration. The divertor has an important influence on the stability, plasma rotation and transport[6]. In the past more than ten years, we have simultaneously realized the low-confinement mode (L-mode) and high-confinement mode (H-mode) discharges, observed the geodesic acoustic mode (GAM) zonal flow[7], the high-energy-trapped-electron-excited fishbone mode[8] and beta-induced Alfven eigenmode (BAE), and the impurity-driven electromagnetic instability[9], and can apply a variety of current drive and auxiliary heating measures to improve the discharge performance. The HL-2A team has carried out various[7–10] studies related to experimental analysis, but the MHD equilibrium and stability of circular and near-circular plasmas need to be improved. The results are helpful to understand Tokamak physics, and have reference application value for optimizing the experimental design and upgrading of the new generation of device such as HL-3.

The tokamak operational beta limit is caused by an instability predicted by ideal MHD theory[11,12]. The main macroscopic instabilities include the kink mode driven by the parallel current and the ballooning mode driven by the pressure gradient. The kink mode is mostly unstable in low mode numbers, while the ballooning mode is mostly unstable in high mode numbers. Ideal low toroidal mode number ($n$) kink and infinite $n$ ballooning modes have been the subject of interest in the design of new experiments and fusion reactors, and are expected to impose limits on the achievable toroidal beta $\beta_t$ in tokamaks[12–14]. Among them, the kink instability is the potentially strongest macroscopic instability. By studying the kink instability, it is expected to find the optimal normalized beta $\beta_N = [\beta_t(\%)/(I_p(MA)/a(m)B_0(T))]$. Todd et al.[13] pointed out that the low $n$ free-boundary kink mode provides the most stringent limit on the operating beta. In the case of plasma separation from the wall, the beta limit of the $n = 1$ kink is equivalent to the ballooning limit[14]. The kink limit of $n = 1$ is identified to be lower than the kink limits of $n = 2$ and $n = 3$. In fact, the low $m/n$ mode is the most important cause of dusrupt, especially the $n = 1$ mode. Therefore, all the results in this paper are based on the study of the $n = 1$ kink instability.

Some of the previous[15,16] studies have dealt with some aspects of MHD instability, but they are not specific to the systematic study of MHD equilibrium and stability. Shen et al. first determined the operational beta limits of the HL-2A tokamak in two discharge modes[17], but their research on MHD stability is less involved. Shen et al.[18] expanded the physical depth and breadth of the above research based on the analysis of experimental data, but they only analyzed the experimental results of divertor discharge. The extended study of MHD equilibrium and stability in limiter discharge experiments is helpful to ensure the completeness of tokamak MHD research. Based on the above reasons, this paper focuses on the circular cross-section plasma of HL-2A limiter discharge, selects the typical discharge to study the plasma equilibrium, and selects the axial safety factors $q_0 = 0.95,\ 1.05$ and $1.3$ to fully cover the internal kink mode of $n = 1, m = 1$ and the external kink mode of $n = 1, m \geqslant 2$. When studying the typical equilibrium and its stability, this paper pays attention to the physical understanding of the observed phenomena, and calculates the operational beta limit, which develops the existing research results. At the same time, the beta limit of non-sawtooth discharge ($q_0 \gg 1$) is calculated, and the result is different from that of sawtooth discharge ($q_0$ varies around 1). The results show some qualitative and quantitative differences between limiter discharge and divertor discharge. Finally, we discuss various cases related to the beta limit of the circular cross-sectional tokamak plamsas.

**2. MHD equilibrium**

The numerical equilibria in this paper are reconstructed or calculated using the EFIT code[19]. The low toroidal mode number (*n*) equilibria are then analyzed for stability using the GATO code[20,21]. The equilibrium[22] are reconstructed by inverting the following Grad-Shafranov equation:

$$\Delta^* \chi \equiv R \frac{\partial}{\partial R}\left(\frac{1}{R} \frac{\partial \chi}{\partial R}\right) + \frac{\partial^2 \chi}{\partial z^2} = -\mu_0 R J_T, \tag{1}$$

$$J_T = RP'(\chi) + \mu_0 \left(F^2(\chi)\right)'/8\pi^2 R. \tag{2}$$

Here, $\chi$ is the poloidal flux per radian of the toroidal angle $\varphi$ enclosed by the flux surface, and $J_T$ is the toroidal current density. Normalized flux is $\psi = (\chi - \chi_0)/(\chi_s - \chi_0)$, where $\chi_0$ and $\chi_s$ are the poloidal flux on the magnetic axis and at the edge, respectively. The two free functions $P'(\psi)$ and $(F^2(\psi))'$ are given by:

$$P'(\psi) = \sum_{n=0}^{n_p} \alpha_n \psi^n, \tag{3}$$

$$\left(F^2(\psi)\right)' = \sum_{m=0}^{n_F} \beta_n \psi^m. \tag{4}$$

Where the polynomial coefficients $\alpha_n$ and $\beta_n$ are given by the specific equilibrium. In practical treatment, we use a current profile of the form:

$$J_T = J_0 \left[\beta_p^0 \frac{R}{R_0} + (1 - \beta_p^0)\frac{R_0}{R}\right](a + b\psi), \tag{5}$$

Where $R_0$ represents the major radius at the center of the plasma; $J_0$ is the current constant; $a$ and $b$ are constant coefficients; $\beta_p^0$ is the initial value of the poloidal beta.

The HL-2A discharges #4206 and #4044 belong to limiter-configuration ohmic discharges. The discharge parameters include plasma current $I_p \approx 300$ kA, electron line average density $\bar{n}_e \sim 1.5 \times 10^{13}$ cm$^{-3}$, toroidal magnetic field $B_T \approx 2$ T. The MHD equilibrium of discharges #4026 is shown in Fig. 1(a), and the other parameters are $\beta_p^0 = 0.8$ and $q_0 = 0.95$. It can be seen that this is a circular cross-section plasma, and its magnetic surface structure is shown in Fig. 1(b). It contains the $q_0 = 1$ magnetic surface. The magnetic surface of $q = 2$ is located at the small radius of plasma 2/3, and $q_{95} = 3.7 - 3.8$.

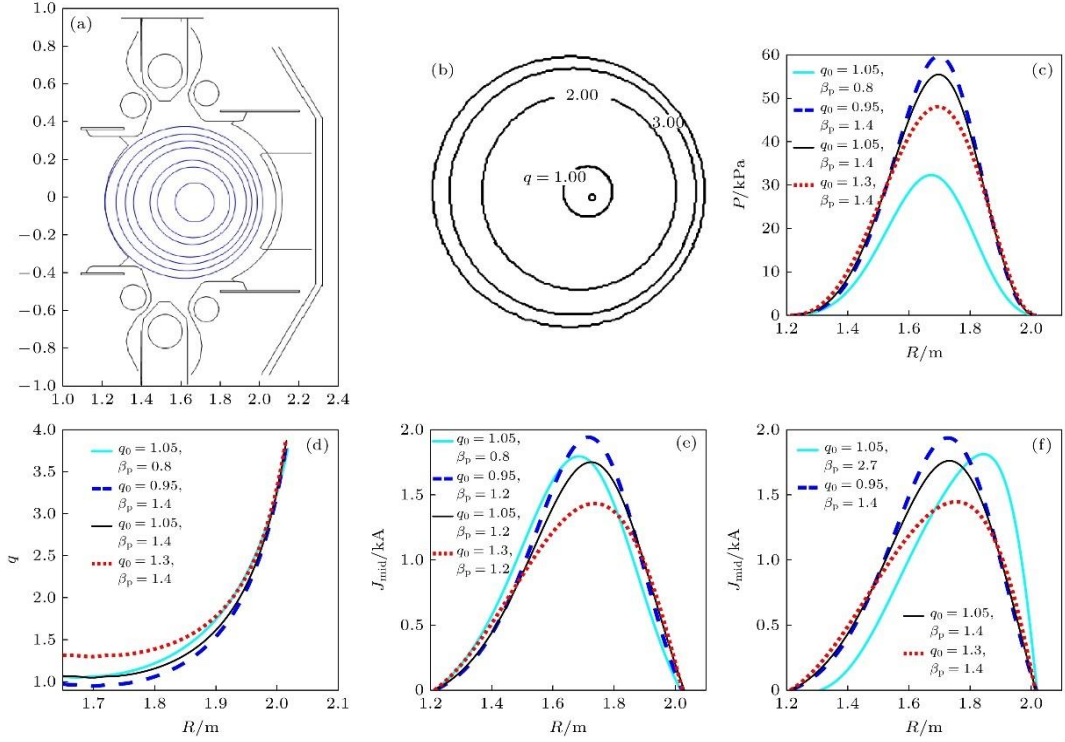

Figure 1. (a) Equilibrium configuration constructed and (b) mapped flux surfaces for $q_0 = 0.95$ and $\beta_p = 0.8$ in HL-2A discharge #4206, and (c) pressure profile, (d) $q$ profile and (e), (f) current density ($J_{\text{mid}}$) profiles for different $q_0$ and $\beta_p$.

The pressure profile, $q$ profile and current profile are given for several sets of parameters in the Fig. 1(c) —(f), respectively. With the change of $q_0$ and $\beta_p^0$, these profiles vary greatly and have strong contrast with each other. The smaller the $\beta_p$ is, the flatter the pressure profile is, the larger the shear in the center of the $q$ profile is, and the closer the current profile is to the standard normal distribution. When the $\beta_p$ increases, the pressure profile becomes sharper, the $q$ profile becomes flatter in the central region, and the current density peaking position moves to the weak field side. When the $\beta_p$ is the same, the change of $q_0$ also has a significant effect on the various profiles. As shown in Fig. 1(c),(d) and(f), when $\beta_p = 1.4$, the pressure and current profiles become flat in turn for $q_0 = 0.95$, 1.05 and 1.3, and the $q$ profile also has a small change. The change of parameters causes the change of profile, which will seriously affect the stability of MHD. For example, the pressure profile is an important factor. When the $\beta_p$ is low, the pressure is wide and flat; When the $\beta_p$ is high, the pressure peaking is steep-as we will see later, the plasma is more stable when the $\beta_\text{p}$ is low, precisely because of the wide flat pressure. In addition, the $q$ profile is also important for MHD stability.

After the equilibrium is obtained by reconstruction and calculation, we use the GATO code to calculate the MHD stability of the equilibrium. GATO uses the finite mixed element method[21] to solve the linearized ideal MHD equations in variational form. The adopted grid number of flux coordinate is $N_\psi \times N_\chi = 100 \times 200$, where $N_\psi$ is the number of flux planes and $N_\chi$ is the number of poloidal angles. Consider a free boundary plasma. The toroidal mode number is set to $n = 1$. The growth rate is normalized to the Alfven frequency $\omega_\text{A} = [B_0^2/\mu_0 \rho R_0^2 q_0^2]^{1/2}$. Where $B_0$ is the toroidal magnetic field at the center of the plasma and $\rho$ represents the mass density. For all growth rates $\gamma$, the equilibrium is considered stable if there are $\gamma^2/\omega_\text{A}^2 \leqslant 10^{-5}$. If there is a $\gamma^2/\omega_\text{A}^2 > 10^{-4}$, the equilibrium is considered unstable. When the growth rate is in the range $10^{-5} < \gamma^2/\omega_\text{A}^2 \leqslant 10^{-4}$, the mode structure needs to be examined to determine stability, that is, the equilibrium with highly localized internal mode structure is considered stable, otherwise the equilibrium is considered unstable. However, if other stability criteria are defined, essentially the same stability result will be obtained as long as it is reasonable.

**3. MHD stability**

3.1 Internal Kink Mode and Axial Safety Factor Effect

When the axial safety factor is $q_0 < 1$, the internal mode of the $m/n = 1/1$ is always unstable; The internal mode can be stabilized by increasing the $q_0$ to be slightly larger than 1. The stability calculation results of $\beta_\text{p} = 1.2, q_0 = 0.95,\ 1.05,\ \text{and}\ 1.3$ are given in Fig. 2. The general feature is that unstable (1,1) internal kink modes (Fig. 2(a) and Fig. 2(b)) appear in the plasma at $q_0 = 0.95$, and the internal kink modes are stabilized at $q_0 = 1.05$, so the plasma is stable (Fig. 2(c) and Fig. 2(d)).

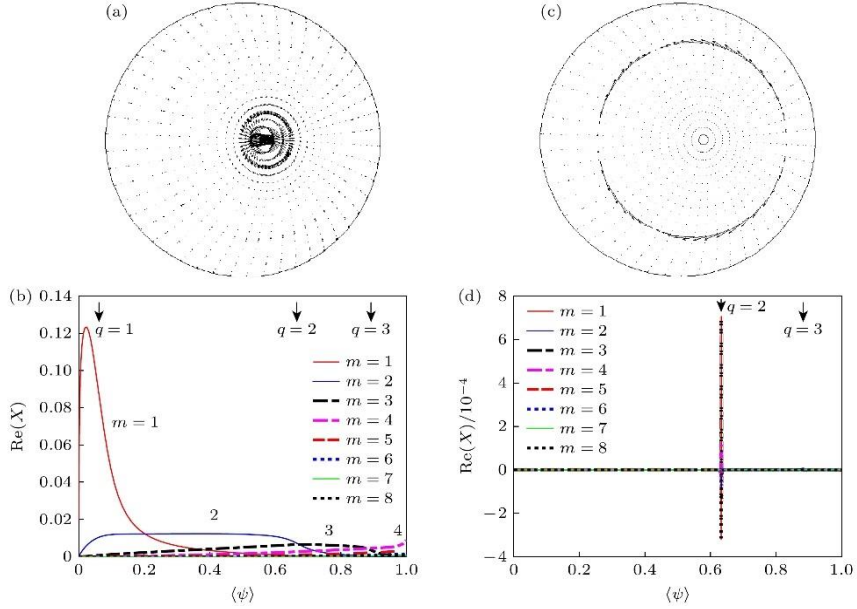

**Figure 2.** At $\beta_p = 1.2$ for discharge #4206, the mode displacement vectors projected onto the poloidal plane (a), (c) and the Fourier decomposition of the normal displacement (b), (d): (a), (b) $q_0 = 0.95$; (c), (d) $q_0 = 1.05$. The horizontal axis $\langle \psi \rangle$ represents the normalized magnetic flux.

Specifically, in a tokamak discharge, unstable internal modes will appear when the safety factor $q$ drops below 1 near the magnetic axis, as shown in the Fig. 2(a). The toroidal mode number of this mode is $n = 1$, and the poloidal mode number is $m = 1$. It is characterized by the interal kink instability, as shown in Fig. 2(b). The $m/n = 1/1$ harmonic of the mode is prominent, and the $m = 2$ (and other higher $m$) harmonics also appear in the Fourier decomposition spectra, that is, in the region between the magnetic axis and the plasma surface, there is coupling of multiple harmonics, but the intensity of the $m \geqslant 2$ harmonics is very weak. The squared linearly normalized growth rate of the instability is $\frac{\gamma^2}{\omega_A^2} = 0.9117 \times 10^{-3}$. In the cylindrical approximation[23], when the axial safety factor $q_0$ is lower than 1, the internal kink mode of $m/n = 1/1$ is essentially unstable at any plasma pressure gradient $(\beta)$[24].

Under the condition of $q_a > 2$ and $q_0 > 1$, the internal kink mode can be stabilized[25]. Fig. 2(c) and Fig. 2(d) show that the internal kink mode is stabilized when the $q_0$ is slightly larger than 1 (here, $q_0 = 1.05$), where $\gamma^2/\omega_A^2 = 0.17 \times 10^{-7}$. In fact, if the pressure gradient (i.e., the poloidal beta $\beta_p$) is small enough, the $n = 1$ internal kink mode can be stabilized by the toroidal effect. On a high aspect ratio tokamak, when $\beta \sim \varepsilon^2$ (where $\varepsilon$ is the inverse aspect ratio), all that is required to achieve the stability of an ideal MHD internal mode for a circular cross-section plasma is that $q > 1$ everywhere in the plasma.

Fig. 3 gives the results of kink mode and its harmonic decomposition of discharge #4044 at $q_0 = 0.95$. When $q_0 = 0.95$, we again see that when $\beta$ is small (e.g., $\beta_p = 0.8$), the internal kink instability dominates, as shown in Fig. 3(a) and Fig. 3(b). When the $\beta$ is very large, such as $\beta_p = 1.8$, the plasma develops into a global instability, and the external kink mode develops strongly, as shown by Fig. 3(c) and Fig. 3(d). For the kink mode, the strongest amplitudes of the internal (kink) mode ($m/n = 1/1$) and the external mode ($m \geqslant 2$) are located at or near the $m = nq$ resonance surface. The external kink mode is perturbed in the whole region from the axis to the plasma boundary. $q_0 \gtrsim 1$ can stabilize the internal kink mode. However, it is worth noting that the fact that the equilibrium with $q_0 < 1$ is unstable is not absolute. Because Kerner et al.[26] have shown that for an ideal kink mode, although it is always unstable in the cylindrical configuration when the axial safety factor is $q_0 < 1$, it can be stable in the toroidal geometry in some cases when $q_0$ takes any value, provided that the poloidal beta $\beta_p$ is lower than some small but critical value.

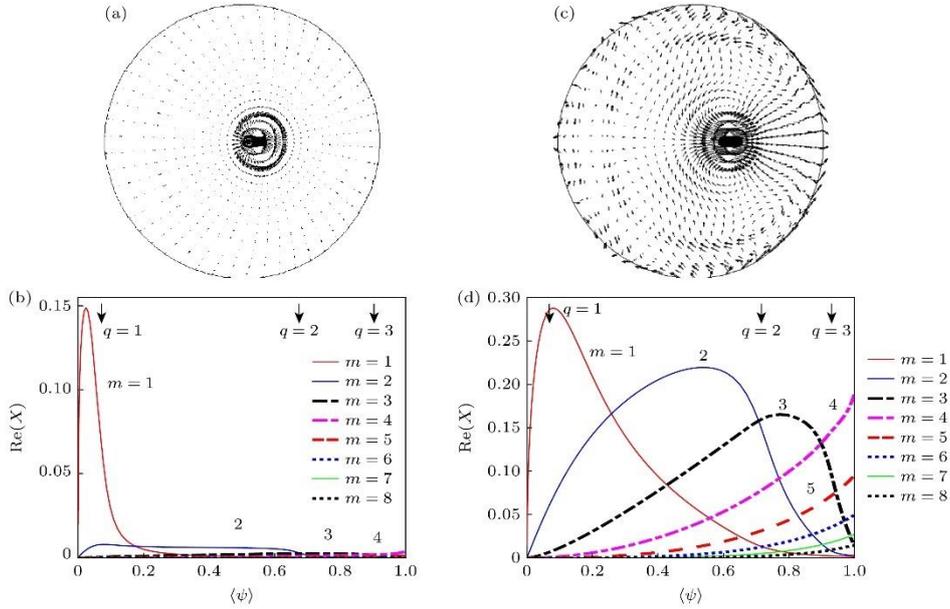

**Figure 3.** Unstable kink mode for limiter discharge with $q_0 = 0.95$ for discharge #4044, the mode displacement vectors projected onto the poloidal plane (a), (c) and Fourier decomposition of the normal displacement (b), (d): (a), (b) $\beta_p = 0.8$; (c), (d) $\beta_p = 1.8$

3.2 External kink mode and beta effect

When the $q_0$ is slightly greater than 1, the plasma will change from stable to unstable with the increase in beta, as shown in Fig. 4. In discharge #4044, the plasma is stable at $q_0 = 1.05$ and $\beta_p = 0.8$, and the eigenvalue of stability is $\gamma^2/\omega_A^2 = 0.1467 \times 10^{-7}$. When the $\beta_p$ increases to 1.8, the external kink mode of global instability appears, the eigenvalue is $0.1336 \times 10^{-2}$, and the intensity of each $m$ harmonic is very high.

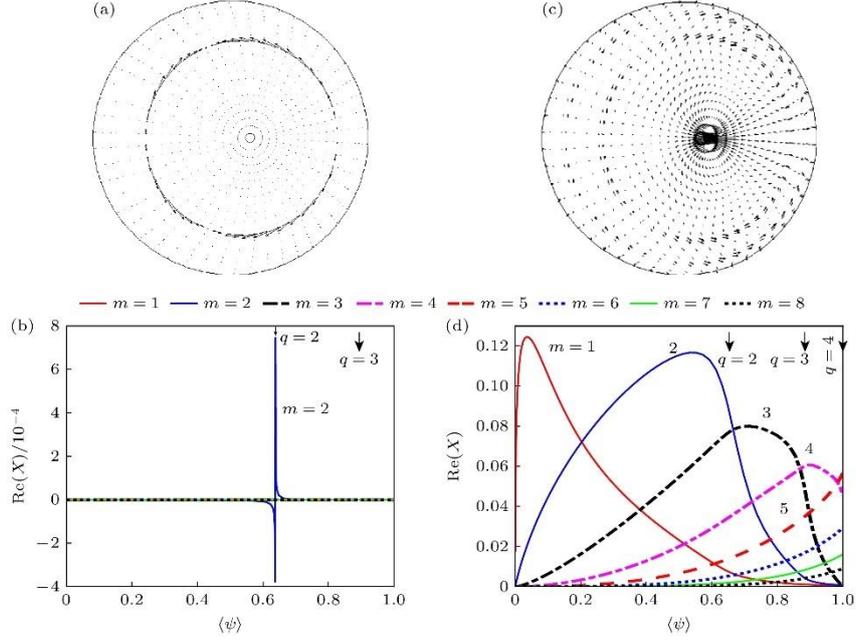

**Figure 4.** For discharge #4044, mode displacement vectors projected onto the poloidal plane (a), (c) with $q_0 = 1.05$ and Fourier decomposition of the normal displacement (b), (d): (a), (b) $\beta_p = 0.8$; (c), (d) $\beta_p = 1.8$.

When the beta is high enough, the external kink mode is dominant in the plasma. Fig. 5 shows the kink modes at different $q_0$ for $\beta_p = 2$. With the increase in poloidal pressure $\beta_p$, the plasma instability develops stronger and stronger, the plasma elongates transversely, and the Shafranov shift increases. Fig. 5(a) and Fig. 5(b) again show that, in general, when $q_0 < 1$ and the $q = 1$ magnetic surface is located within the plasma, the unstable mode is dominated by the $m = 1$ Fourier harmonic. When $\beta_p = 2$ is exceeded, the calculation shows that the growth rate (Fig. 5(a)) and the edge amplitude (Fig. 5(b)) increase sharply. The rapid growth of the edge amplitude is an indicator of the change in the unstable mode signature from an initial internal kink to an external kink with a large destructive force. Fig. 5(c) and Fig. 5(d) show that this mode also has a large contribution from several $m > 1$ harmonics when $q_0 > 1$. When $q_0 = 1.05$, there is an overlap between the two types, especially at high $\beta_p$. Although at low $\beta_p$, making $q_0 > 1$ can stabilize the kink mode. However, when $\beta_p = 2$, the plasma is globally unstable in all $q_0$ cases, and the harmonic range is very wide, including $m = 1,2,3,\cdots$ and other branches. When $q_0 = 1.3$, the plasma instability growth rate further increases ($\approx 0.02$) (Fig. 5(e)), and the oscillation amplitude of each $m$ harmonic (including $m = 1$ harmonic) is very large (Fig. 5(f)). In summary, when $q_0 > 1$, if $\beta_p$ is large enough, for example, here the $\beta_p = 2, m/n = 1/1$ branch still appears in the Fourier decomposition diagram, because the internal kink mode becomes unstable again when $\beta_p$ exceeds a certain finite threshold. However, in the case of high $\beta_p$, the larger the $q_0$ is, the more the MHD instability is dominated by the external kink mode, and the smaller the contribution of the internal kink mode to the instability is.

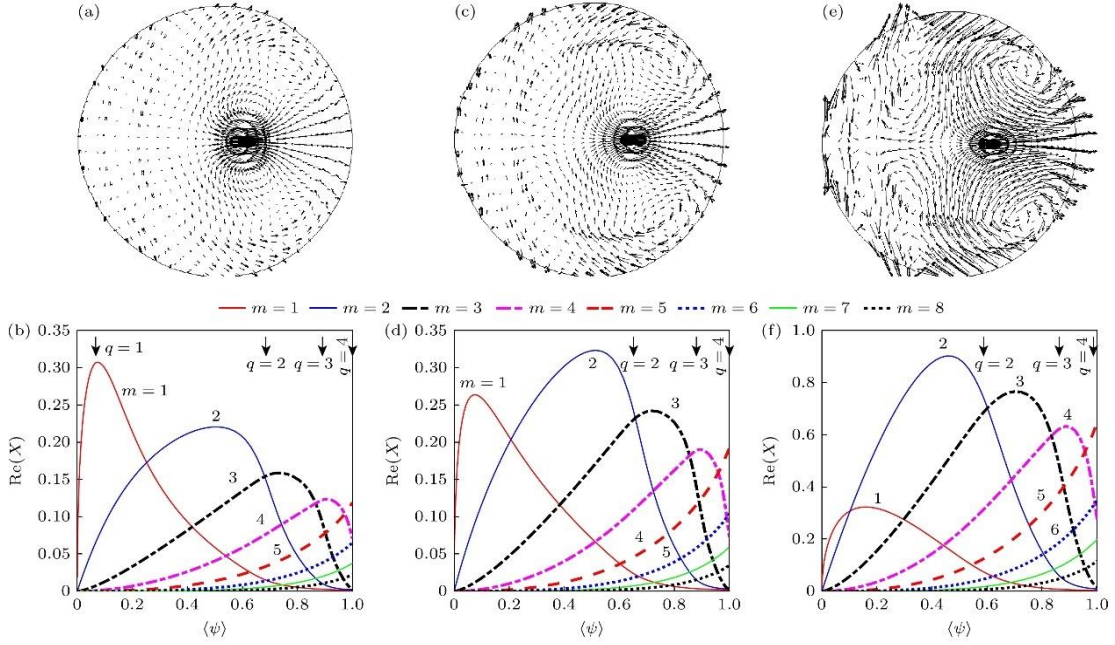

**Figure 5.** At $\beta_p = 2$ for discharge #4206, the mode displacement vectors projected onto the poloidal plane for $q_0 = 0.95$ (a), $1.05$ (c) and $1.3$ (e), and Fourier decomposition of the normal displacement for $q_0 = 0.95$ (b), $1.05$ (d) and $1.3$ (f).

Fig. 5 shows that outside the $q = 2$ magnetic surface, the $m = 1$ harmonics are coupled to the $m = 2$ (and higher $m$) kink harmonics. For the internal kink, the $m = 1$ harmonic perturbation disappears at the boundary and does not contribute significantly to the total perturbation potential energy. However, the kink harmonics of $m = 2$ and higher $m$ are still strong at the boundary, especially the harmonics of $m \geqslant 5$ are the strongest at the boundary. It is worth pointing out that the most important of the external kink modes is the $2/1$ mode. This mode can develop to a large scale, causing the rotation to slow down and eventually lock up, causing a disrupt. It limits the maximum beta that a tokamak can achieve, known as the hard beta limit. On the other hand, combining the results given in Fig. 1 and Fig. 2-Fig. 4, it can be found that the pressure gradient is destabilizing for the kink mode and the magnetic shear is stabilizing. At the same time, comparing the current profile with the stability results at the same $q_0$, it is found that the sharp current profile is beneficial to reduce the growth rate. In summary, it is demonstrated here that the stability of the $n = 1$ ideal kink is heavily dependent on the current and pressure profiles.

Interestingly, from Fig. 4(d) as well as Figs. 5(b),(d) and (f), it can be seen that when $m < q_a$, the intensity of the $m$ subharmonic at the plasma surface ($\psi = 1$) becomes 0., that is, when $m < q_a$, the surface mode is stable. When $m > q_a$, the intensity of the $m$-th harmonic is the largest at the surface. That is to say, the plasma surface disturbance mainly comes from the harmonic of $m > q_a$. On the other hand, the $q = 2$ magnetic surfaces studied here are all located at or near the $0.6a$. From the numerical calculation, the true internal kink growth rate is determined by a fine balance of contributions from the inside and outside of the $q = 1$ surface. These two mode types cannot be clearly distinguished from linear stability calculations for a single discharge equilibrium. Because when $q_0 < 1$, the unstable mode almost always contains some mixture of internal and external components, accompanied by the largest component $m = 1$ harmonic, as shown by Fig. 2 and Fig. 3.

3.3 Stability interval and marginal safety factor

The combined $(q_0, q_a)$ of the axial and edge safety factors determines the plasma equilibrium with a parabolic safety factor profile[18]. The stability of the equilibrium is determined jointly by $q$ and the $\beta$. To analyze the average effect of plasma beta, it is necessary to introduce the root mean square (RMS) beta $\beta^*$:

$$\beta^* = 2\mu_0 \langle p^2 \rangle^{1/2} / B_t^2. \tag{6}$$

Shen et al.[18] analyzed the physical mechanism of plasma stability determined by the combination of $q_a$ and $\beta$. Here we discuss the stability of the plasma in the $1/q_a$-$\beta^*$ parameter plane. Fig. 6(a) and Fig. 6(b) show that at a fixed $q_0 = 1.05$, the overall trend of the RMS beta value $\beta^*$ of the equilibrium constructed by the $q_0, q_a$ combination with the $1/q_a$ diagram is that the smaller the $1/q_a$ is, the larger the $\beta^*$ is. For both $n = 1$ and $n = 2$ modes to be stable, $q_a$ is required to be low and $0.4 \lesssim \beta^* < 1.25$; for both modes to be unstable, $q_a$ is required to be high and $\beta^* > 1.5$; and for $n = 1$ mode to be stable but $n = 2$ mode to be unstable, the corresponding $q_a$ value is necessary to be medium, and the $\beta^*$ is between 1.25 and 1.5. That is to say, when $1/q_a$ is large and $\beta_p$ is small, it is easier for both $n = 1$ and $n = 2$ kink modes to become stable. On the contrary, the region where both kink modes are unstable is characterized by a small $1/q_a$ and a large $\beta_p$. In the intermediate region between the above two cases, the $n = 1$ mode is stable and the $n = 2$ mode is unstable. Fig. 6(b) and Fig. 6(c) show the variation of $1/q_a$ from the maximum $\beta^*$ defined by the stabilization of the $n = 1$ and $n = 2$ kink modes, respectively, without fixing $q_0$. The maximum $\beta^*$ required for the stability of $n = 1$ mode decreases with the increase of $1/q_a$, but after $1/q_a > 0.275$, $\beta^*$ increases with the increase of $1/q_a$. For the $n = 2$ mode, the maximum $\beta^*$ decreases monotonically with the increase of $1/q_a$, or the maximum $\beta^*$ increases with the increase of $q_a$.

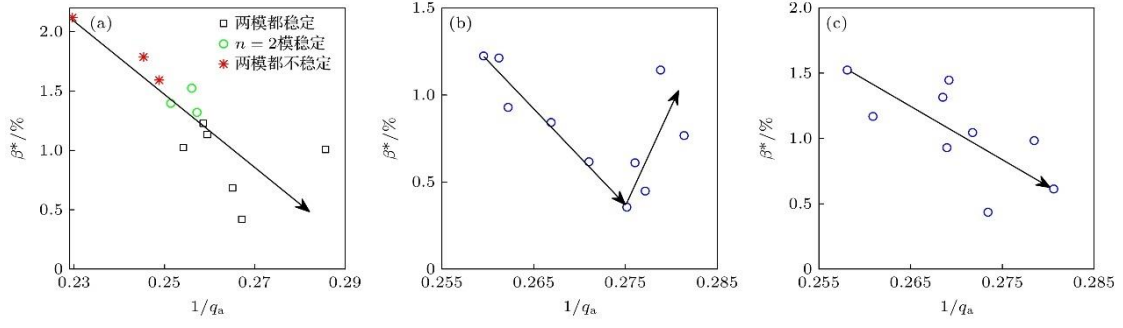

**Figure 6.** For the typical limiter discharges: (a) Kink stabilities in $1/q_\mathrm{a}$-$\beta^*$ plane at fixed $q_0 = 1.05$; (b) $\beta^*$ vs. $q_\mathrm{a}$ at unfixed $q_0$, here $\beta^*$ is the maximum achievable one limited by $n = 1$ kink; (c) $\beta^*$ vs. $q_\mathrm{a}$ at unfixed $q_0$, here $\beta^*$ is the maximum achievable one limited by $n = 2$ kink. The solid lines with arrows indicate the change direction of $\beta^*$ as $q_\mathrm{a}$ increases.

## 4. Beta limit

The external kink limits the maximum beta that the device can achieve. We calculate the operational beta limit of the device based on discharge #4044 and find $\beta_\mathrm{N}^\mathrm{c} = 2.01$. Next, we focus on the physical basis of the calculation. In this calculation, the dependence of the square of the normalized growth rate on $\beta_\mathrm{p}$ and $\beta_\mathrm{N}$ is shown in Fig. 7(a) and Fig. 7(b), respectively. The variation of the maximum edge disturbance displacement with the normalized beta $\beta_\mathrm{N}$ is given in Fig. 7(c). First, the results of $q_0 \sim 1$ ($q_0 = 0.95$ and $q_0 = 1.05$) are analyzed. Experimentally, $q_0 > 1$ or $q_0 < 1$ can be distinguished by sawtooth behavior. From Fig. 7(a) and Fig. 7(b), it can be seen that $\beta_\mathrm{p} = 1.36$ or the corresponding $\beta_\mathrm{N}^\mathrm{c} = 2.01$ is the critical point of the dependence of the normalized growth rate on $\beta_\mathrm{p}$ and $\beta_\mathrm{N}$, respectively. Below this critical point, the plasma instability growth rate is very low at $q_0 = 0.95$, and the normalized squared growth rate at $q_0 = 1.05$ is $\gamma^2/\omega_A^2 < 10^{-4}$. Above this critical point, the growth rates of both $q_0 = 0.95$ and $q_0 = 1.05$ begin to increase rapidly. As shown in Fig. 7(c), the variation of the maximum disturbance displacement is similar to that of the growth rate. When $\beta_\mathrm{N} < 2.01$, the perturbation amplitude of $q_0 = 0.95$ mode is much higher than that of $q_0 = 1.05$. However, when $\beta_\mathrm{N} > 2.01$ and $q_0 > 1$, the disturbance of the mode increases sharply, and the disturbance amplitude is much higher than that of $q_0 = 0.95$. This critical $\beta_\mathrm{N}$ (about 2.0) is the operational beta limit of the sawtooth discharge with circular cross-section configuration. In the case of $q_0 = 1.3$, the critical point is smaller, and the critical poloidal beta and the critical normalized beta are $\beta_\mathrm{p}^\mathrm{c} = 1.20$ and $\beta_\mathrm{N}^\mathrm{c} = 1.81$.

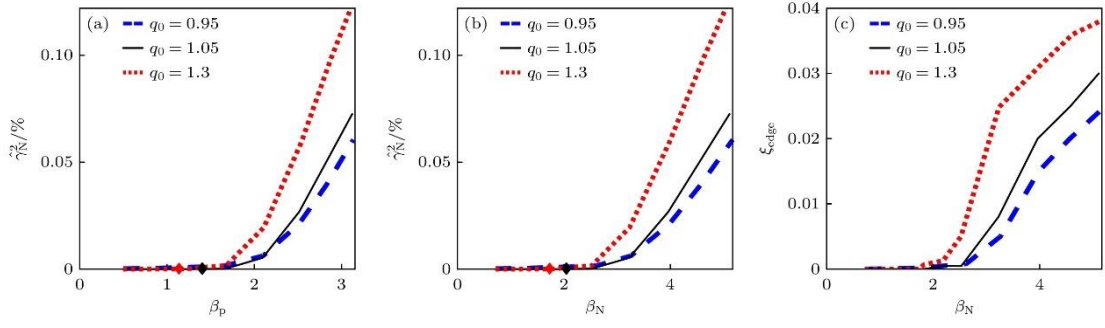

**Figure 7.** Calculations were based on the data of discharge #4044: (a), (b) Square value of normalized mode growth rate $\hat{\gamma}_N^2$ as functions of $\beta_p$ and $\beta_N$; (c) maximum edge normal displacement as functions of normalized. Note that $\beta_p$ is the actual polar beta of the calculated equilibrium configuration. The rhombus symbol represents the critical point

There are two points worth noting here. First, when $q_0$ further increases to 1.3, the beta limit decreases, that is, $\beta_N^c = 1.81$. When $\beta_N$ exceeds this value, the edge disturbance amplitude of the mode at $q_0 = 1.3$ is much higher than that at $q_0 = 0.95$ and $q_0 = 1.05$. This is evidence that too high a $q_0$ is detrimental to MHD stability. Secondly, when $\beta_N$ grows beyond $\beta_N^c$, the calculation of the unstable mode at $q_0 < 1$ is essentially still determined by the composition of the (1,1) mode inside the $q=1$ magnetic surface, but the composition of the harmonic outside the $q=1$ magnetic surface and the $m \geqslant 2$ harmonic show greater contributions to the mode instability, and their growth rates begin to accelerate and quickly approach the growth rates of the corresponding harmonics at $q_0 > 1$. The critical $\beta_N$ associated with this transition can be described as the "operational $\beta$ limit" at $q_0 < 1$. In addition, it should be noted that the discharge with parabolic current profile mainly operates in the adjacent areas on the left and right sides of $q_0 = 1$, so the beta limit $\beta_N^c \cong 2.0$ obtained at $q_0 \gtrsim 1$ is the most representative. However, the case of "$q_0 \gg 1$" mainly occurs in the discharge with reverse magnetic shear configuration, so the beta limit $\beta_N^c \cong 1.8$ at $q_0 = 1.3$ obtained from the study of parabolic current profile discharge here is not generally representative in the experiment.

## 5. Discussion

The semi-empirical scaling law[27] for the tokamak beta limit can be approximately written as:

$$\beta(\max) = C_\beta (I/aB_0). \tag{7}$$

Here the value of the coefficient $C_\beta$ depends on the MHD instability. Troyon and Gruber derived[28] the maximum mean beta limited by ideal MHD instability for tokamaks with different shapes and aspect ratios as $\beta < (2.0 - 2.5)(I/aB_0)$. Based on the study of circular cross-section plasma in the HL-2A tokamak, the value of the coefficient $C_\beta$ corresponding to the $q_0$ slightly greater than 1 is 2.01, which is basically consistent with the lower limit of the classical theory. For the $q_0 < 1$ case, there is no beta limit, and the plasma is MHD unstable in all cases, but it can cause disruption only at large poloidal betas. When $q_0 > 1$, the MHD instability imposes an operational beta limit on the tokamak due to the stabilization of the internal kink mode. The $C_\beta$ at $q_0 = 1.05$ is often referred to as the maximum operational beta $\beta_N^c$. The pressure and current density profiles of the single-parameter family were optimized[28] based on the $n = 1$ kink mode and the high $n$ ballooning mode, resulting in $C_\beta = 2.8$ (in% ·m ·T ·MA$^{-1}$). If the kink mode is optimized separately, one gets $C_\beta = 3.2$. Bondeson et al.[29] further optimized the pressure profile to be critically stable everywhere for high $n$ ballooning modes, and obtained a $C_\beta$ value of 3.7—4.4. There are some differences between the results of this paper and those of profile optimization, which is a meaningful topic and is discussed as follows.

The maximum beta that can be achieved in the device depends strongly on the pressure and current profiles, and on the shape of the profiles[12]. For a wide pressure profile and a sharp current density profile, the beta limit increases. Because the external kink instability is caused by a strong current gradient close to the plasma surface. The wider pressure profile and reduced shear are beneficial to limiting the internal kink instability. Thus, within the framework of ideal MHD, there are three directions that can be taken to increase the beta limit: changing the current and pressure profiles and the stabilization of the wall.

Shen et al.[30] studied the stabilization effect of a perfectly conductive wall on kink modes based on the HL-2A tokamak experiment, and pointed out that the wall stabilization is a method to suppress the stability of low *n*. The ideal conductive wall can partially suppress the surface kink component, so that the operational beta limit of the device is directly increased by about 6.5%. If the active feedback control[31] is considered at the same time, the operational beta is expected to be greatly improved. In addition, the wall is very effective for stabilizing axisymmetric MHD modes of tokamak plasmas with high elongation. Of course, while the kink mode can be stabilized by a conductive wall surrounding the plasma, the finite resistivity of the wall leads to another instability called resistive wall mode (RWM). Resistive wall mode can also be regarded as a special form of external kink mode, but the growth rate of RWM is much slower than that of ideal kink-ballooning mode, so the study of resistive wall mode is another hot spot[32,33].

It should be pointed out that although ideal MHD theory is expected to give an upper limit of stability, the experimental plasma cannot be fully described by ideal MHD, which treats the plasma as a perfectly conducting fluid. The pressure-driven resistive instability plays an important role in limiting the experimental pressure. Some current profiles optimized for ballooning and kink modes may be unstable to tearing modes[34]. Therefore, due to the increase and superposition of practical limiting conditions, the experimental operational beta limit is more complex than the ideal operational beta limit, and sometimes smaller. However, the operational beta may exceed the ideal MHD operational beta limit when the plasma cross-section is reasonably optimized (mainly the optimization of elongation ratio and triangular deformation) and the pressure and current profiles are ideal[35]. In addition, the beta limit is also affected by high-energy particle effects.

In this paper, an unoptimized circular plasma is studied. In fact, the optimization of section shape is particularly beneficial to improve the operational beta limit. Strong cross-sectional shape is the key to solving the conflicting requirements of $n = 0$ axisymmetric stability, $n = 1$ mode and ballooning stability. Deformation, including increased elongation ratio $\kappa$, triangular deformation $\delta$, etc., has a complex effect on kink boundaries. Low aspect ratios tend to provide the same stability benefits as high $\kappa$ and strong $\delta$ because it has a similar effect on extending field line lengths in regions of good curvature and shortening correlation lengths in regions of bad curvature. A lower aspect ratio generally widens the stability region. In the upgrading of HL-2A, the new device HL-3 has made great changes in the cross-section shape, one of which is to change the circular and near-circular cross section of HL-2A into an inverse triangular cross section, and to consider the effect of the wall.

## 6. Summary

The plasma beta is limited by MHD instability, and the theoretical beta limit is determined by the development of the instability. In this paper, the MHD equilibrium and stability of circular tokamak plasma are studied, and then the beta limit of the device is studied, which can provide guidance for the optimal design and high performanced operation of tokamak. To sum up, the following results can be obtained from the research in this paper.

1) For large aspect ratios, a stable equilibrium of the $\beta_\mathrm{p} \sim 1$ can be generated with a reasonable current profile. The key safety factor profile parameters affecting stability are $q_0$. The $q_0 \gtrsim 1$ can stabilize the internal kink mode at low $\beta_\mathrm{p}$. However, higher $q_0$ is not beneficial for stability, although other studies have shown that high $q_0$ reversed-magnetic-shear configurations can still have some benefits for confinement (see Bondeson et al.[29]). The external kink mode is global and the internal kink mode is local. For the $n = 1$ external kink, this mode is more restrictive than the other low $n$ modes when it is assumed that there is no external-conducting-wall stabilization[14]. Therefore, the operational beta limit study is limited to the $n = 1$ kink limit. By properly selecting the current distribution, it is found that a flat distribution and a low value of the poloidal beta $\beta_p$ are the most favorable. However, the flat current distribution causes a strong external kink instability. The peak current distribution can suppress these instabilities at the expense of a reduction in the value of the specific voltage. When the $\beta_p$ is high, the external kink mode is fully developed, which limits the maximum that the device can reach $\beta_t$. When the $\beta_p$ is higher, the internal kink can be seen to be coupled by these higher *m* external kink modes.

2) The maximum beta limit imposed on the plasma by the kink mode instability is called the hard beta limit. In this paper, the upper limit of the beta limit in the HL-2A tokamak discharge is predicted by using the ideal MHD theory, which deepens the research results of Shen et al.[17]. Taken together, the operational beta limit corresponding to $q_0 = 1.05$ is proportional to the normalized current $I_N (= I/aB)$, i.e., $\beta_N^c = 2.01 I/aB$.

Significant agreement between the optimized beta limits and the achievable beta value in some discharges was found in some cases[25]. The configuration studied in this paper has not been optimized for shape and profile. At the same time, the object of study is to select the discharge among ohmic or low power auxiliary heating, low confinement (L mode) ones. Therefore, the operational beta limit obtained is not high, but it reaches the lower limit of the theoretically operational beta limit of circular cross-section tokamak. It should also be noted that the boundary pressure gradient and the average edge current density on the surface are limited to 0 in this paper. However, in fact, the presence of a finite pressure gradient at the boundary can lead to an increase in the value of $\beta_N$. For boundary-kink modes, the edge current does not always provide a destabilizing effect; With the increase of the edge current, the linear growth rate increases first and then decreases[36]. All these provide clues for increasing the operational beta. This needs to be taken into account when studying how the Huanliuqi device achieves a high $\beta_N$ discharge.

1700